\documentclass[prd,twocolumn,nopacs,floatfix,amsmath,nofootinbib,amssymb,floatfix]{revtex4}
\usepackage{graphicx,color,dcolumn,booktabs,bm}
\usepackage{longtable,lscape}
\usepackage{txfonts}
\usepackage{overpic}
\usepackage{amssymb}
\usepackage{indentfirst}
\usepackage{feynmf}   
\usepackage{slashed}  
\usepackage{cases}
\usepackage{color}
\usepackage{multirow}
\usepackage{epstopdf}
\usepackage{enumerate}
\usepackage{graphicx,color,dcolumn,booktabs,bm}
\usepackage[colorlinks, citecolor=blue,anchorcolor=red,menucolor=red, linkcolor=red,filecolor=red,urlcolor=blue,frenchlinks=red]{hyperref}

\graphicspath{{Figures/}} %
\begin{document}

\title{Charmonium decays into ${\Lambda}_c\bar{\Lambda}_c$ pair governed by the hadronic loop mechanism}
\author{Ri-Qing Qian$^{1,2}$}\email{qianrq18@lzu.edu.cn}
\author{Jun-Zhang Wang$^{1,2}$}\email{wangjzh2012@lzu.edu.cn}
\author{Xiang Liu$^{1,2,3}$\footnote{Corresponding author}}\email{xiangliu@lzu.edu.cn}
\author{Takayuki Matsuki$^4$}\email{matsuki@tokyo-kasei.ac.jp}
\affiliation{$^1$School of Physical Science and Technology, Lanzhou University, Lanzhou 730000, China\\
$^2$Research Center for Hadron and CSR Physics, Lanzhou University $\&$ Institute of Modern Physics of CAS, Lanzhou 730000, China\\
$^3$Lanzhou Center for Theoretical Physics, Key Laboratory of Theoretical Physics of Gansu Province,
and Frontiers Science Center for Rare Isotopes, Lanzhou University, Lanzhou 730000, China\\
$^4$Tokyo Kasei University, 1-18-1 Kaga, Itabashi, Tokyo 173-8602, Japan}

\date{\today}

\begin{abstract}
In this work, we investigate the open-charm decay process $\psi\to\Lambda_c\bar{\Lambda}_c$ via the hadronic loop mechanism for vector charmonia above $\Lambda_c\bar{\Lambda}_c$ threshold. The branching ratios of these vector charmonium states to $\Lambda_c\bar{\Lambda}_c$ are estimated. The charmonium explanation of the $Y(4630)$ observed in $e^+e^- \to \Lambda_c\bar{\Lambda}_c$ is tested. Furthermore, for the predicted higher vector charmonia above 4.7 GeV, the branching ratios $\mathcal{B}[\psi(nS)\to\Lambda_c\bar{\Lambda}_c]$ with $n=7,8,9$ are found to be of the order of magnitude of $10^{-4}-10^{-3}$ while $\mathcal{B}[\psi(mD)\to\Lambda_c\bar{\Lambda}_c]$ with $m=6,7,8$ are of the order of magnitude of $10^{-3}-10^{-2}$. The experimental signals of these missing charmonium states are discussed. The search for them may be an interesting topic in the future BESIII and Belle II experiments.
\end{abstract}

\maketitle

\section{\label{sec1}Introduction}
As the important member of hadron spectroscopy, charmonium family has attracted extensive attention from both theorist and experimentalist since the first charmonium $J/\psi$ was discovered in 1974 \cite{Aubert:1974js,Augustin:1974xw}. Since charmonium is a typical mesonic system composed of charm and anti-charm quarks, charmonium corresponding to low-energy particle physics can be treated as ideal platform to help us to deepen our understanding of non-perturbative behavior of quantum chromodynamics (QCD), which is full of challenge and opportunity.

\begin{figure}[htbp]
  \centering
  \includegraphics[width=8cm,keepaspectratio]{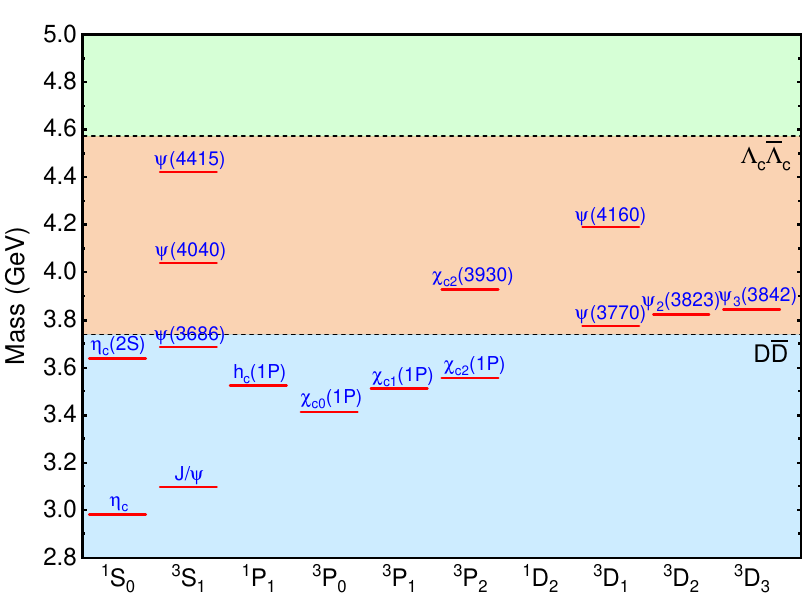}
  \caption{ The established charmonium states and comparison with $D\bar{D}$ and $\Lambda_c\bar{\Lambda}_c$ thresholds. }\label{fig:spectrum}
\end{figure}

At present, the number of charmonium states reported in experiment is constantly increasing, especially, with the observation of a series of charmoniumlike $XYZ$ states in the past 18 years \cite{Liu:2019zoy,Brambilla:2019esw}. In Fig. \ref{fig:spectrum}, we summarize the present status of the charmonium family. We may find that there exist 8 charmonium states below the $D\bar{D}$ threshold, which results in the narrow widths of these charmonia due to the Okubo-Zweig-Iizuka (OZI) rule. Above the $D\bar{D}$ threshold, more charmonia can be found, where the OZI-allowed decay channels composed of charmed and anti-charmed mesons have dominant contribution to the corresponding total width of these involved charmonia.
For these charmonia above the $\Lambda_c\bar{\Lambda}_c$ threshold, a new type of OZI-allowed channel $\Lambda_c\bar{\Lambda}_c$ is open. The $\Lambda_c\bar{\Lambda}_c$ decay channel attracts less attention because all the established charmonia cannot decay into this channel.

The construction of higher charmonia with masses above the $\Lambda_c\bar{\Lambda}_c$ threshold is ongoing, which is stimulated by the observation of the $Y(4660)$ in $e^+e^-\to \psi(2S)\pi^+\pi^-$ \cite{Wang:2014hta} and the $Y(4630)$ in $e^+e^-\to \Lambda_c\bar{\Lambda}_c$ \cite{Pakhlova:2008vn}. The charmonium explanation is always the one of the popular theoretical opinions of $Y(4660)$ and $Y(4630)$ \cite{Guo:2008zg,Ebert:2008kb,Albuquerque:2008up,Albuquerque:2011ix,Chen:2010ze,Zhang:2010mw,Sundu:2018toi,Cotugno:2009ys,Dai:2017fwx,Guo:2010tk,Dubynskiy:2008mq,Lee:2011rka,Cao:2019wwt,Li:2009zu,Ding:2007rg,Wang:2020prx}. In Ref. \cite{Li:2009zu}, Li and Chao indicated that the $Y(4660)$ and $Y(4630)$ are the candidate of $\psi(6S)$. Ding et al. \cite{Ding:2007rg} proposed that the $Y(4660)$ can be the $\psi(5S)$ state. In our former work \cite{Wang:2020prx}, we found that the $Y(4660)$ and $Y(4630)$ may have close relation to two charmonium states $\psi_{6S-5D}'$ and $\psi_{6S-5D}''$, which are the mixture  of $\psi(6S)$ and $\psi(5D)$.
When carrying out these studies, especially considering the $e^+e^-\to \Lambda_c\bar{\Lambda}_c$ process, how to quantitatively calculate the $\Lambda_c\bar{\Lambda}_c$ decay of higher charmonia becomes a key point. However, the study of the strong decay of $\psi\to\Lambda_c\bar{\Lambda}_c$ is not enough.

The OZI-allowed strong decay of charmonium to $D\bar{D}$ was well described by the quark pair creation (QPC) model \cite{Micu:1968mk,LeYaouanc:1977gm} which assumes the creation of one light quark pair with vacuum quantum number $J^{PC}=0^{++}$. However, the $\psi\to\Lambda_c\bar{\Lambda}_c$ needs the creation of two light quark pairs, which is beyond the naive QPC model. Xiao {\it et al.} \cite{Xiao:2018iez} chose to extend the QPC model by directly assuming the same strength of two quark-antiquark pair creation vertexes. Simonov \cite{Simonov:2011jc,Simonov:2011cm} developed a double string breaking model with $\Lambda_c\bar{\Lambda}_c$ emission to depict this process. These models are tentative explorations to the mechanism of the $\psi\to\Lambda_c\bar{\Lambda}_c$ process.

In this work, we apply the hadronic loop mechanism, which has been widely applied to the explorations of hidden-charm and hidden-bottom decays of charmonium and bottomonium
\cite{Liu:2006dq,Liu:2009dr,Zhang:2009kr,Meng:2007tk,Meng:2008dd,Meng:2008bq,Chen:2011qx,Chen:2011zv,Chen:2011pv,Chen:2014ccr,Wang:2016qmz,Huang:2017kkg,Huang:2018pmk,Huang:2018cco}, to study the $\psi\to\Lambda_c\bar{\Lambda}_c$ decays. Here, these higher charmonia decay into $\Lambda_c\bar{\Lambda}_c$ occur via the intermediate hadronic loops composed of charmed meson, anti-charmed meson and nucleon. Later, we will present the details of these discussed decay processes.

As an available application of the hadronic loop mechanism in $\psi\to\Lambda_c\bar{\Lambda}_c$, we examine the branching ratios of the $\Lambda_c\bar{\Lambda}_c$ decays of charmonia $\psi_{6S-5D}'$ and $\psi_{6S-5D}''$ proposed to explain the $Y(4630)$ in our previous work \cite{Wang:2020prx}. We find that the branching ratios of $\psi_{6S-5D}'\to\Lambda_c\bar{\Lambda}_c$ and $\psi_{6S-5D}''\to\Lambda_c\bar{\Lambda}_c$ extracted from the $e^+e^-\to\Lambda_c\bar{\Lambda}_c$ data are well reproduced within the hadronic loop mechanism.
Furthermore, we also estimate the $\Lambda_c\bar{\Lambda}_c$ branching ratios of more higher charmonia predicted in Ref. \cite{Wang:2020prx}. In order to arouse the interest of experimentalists, we try to reproduce the rough data of $e^+e^-\to \Lambda_c\bar{\Lambda}_c$ up to center-of-mass (CM) energy of 4.9 GeV based on our theoretical results of branching ratios. By this study, we want to show the possible evidence of higher vector charmonia with masses above 4.7 GeV existing in the present experimental data of the $e^+e^-\to \Lambda_c\bar{\Lambda}_c$ process.

This paper is organized as follow. Firstly, we introduce a hadronic loop mechanism for $\psi\to\Lambda_c\bar{\Lambda}_c$ in Sec. \ref{sec2}. Then the applications of the hadronic loop mechanism to the decay of $\psi_{6S-5D}'$, $\psi_{6S-5D}''$ and higher charmonia were presented in Sec. \ref{sec3}. Finally, we conclude this paper in Sec. \ref{sec4}, where possible signals of higher charmonia above 4.7 GeV existing in the present experimental data are discussed.

\section{\label{sec2} The hadronic loop mechanism in decay process of $\psi \to \Lambda_c \bar{\Lambda}_c$}
\begin{figure}[htbp]
  \centering
  \includegraphics[width=8cm,keepaspectratio]{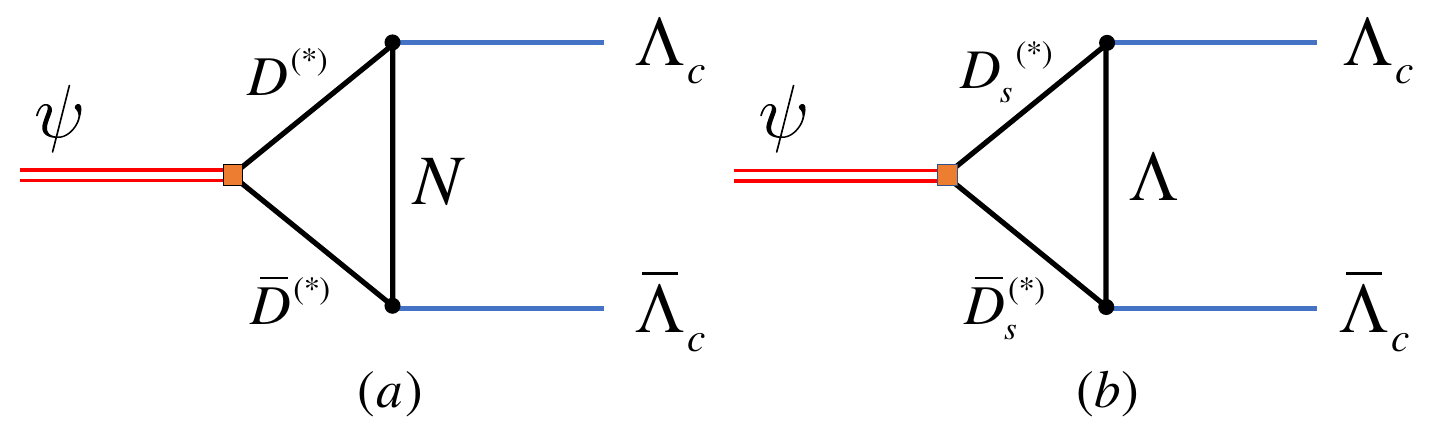}
  \caption{ The schematic diagrams of a hadronic loop mechanism that contribute to $\psi\to\Lambda_c\bar{\Lambda}_c$.}\label{fig:triangle}
\end{figure}

The dominant decay channels of a charmonium states above charmed meson pair threshold are usually the two-body open-charm modes composed of charmed meson pairs. The decay of higher charmonia to $\Lambda_c\bar{\Lambda}_c$ needs one more light quark-antiquark pair creations than the decay channel composed of charmed meson pairs.
Here, we suppose the decay process $\psi\to\Lambda_c\bar{\Lambda}_c$ can proceed via a two-step way. Firstly, the $\psi$ decays into their dominant final states $D^{(*)}_{(s)}\bar{D}^{(*)}_{(s)}$, and then the $D^{(*)}_{(s)}\bar{D}^{(*)}_{(s)}$ is transformed to final states $\Lambda_c\bar{\Lambda}_c$ via exchanging a baryon. This is the hadronic loop mechanism existing in the decay process $\psi\to\Lambda_c\bar{\Lambda}_c$. The two possible Feynman diagrams are shown in Fig. \ref{fig:triangle}. Here, the intermediate charmed meson $D^{(*)}$ in Fig. \ref{fig:triangle} can be either $D^{(*)0}$ or $D^{(*)+}$.

To evaluate these processes, we adopt the effective Lagrangian approach. We need the Lagrangian depicting the interaction between a charmonium state and charmed/charmed-strange meson pairs and the Lagrangian of the couplings involving in charmed baryon $\Lambda_c$ and a charmed/charmed-strange meson together with a baryon $N$/$\Lambda$.
Using the forms listed in Ref. \cite{Huang:2018cco}, we can obtain the Lagrangian of the $\psi \mathcal{D^{(*)}D^{(*)}}$ interaction, which respects the heavy quark symmetry \cite{Casalbuoni:1996pg} and reads as

 \begin{equation}\label{eq:lagrangians}
   \begin{split}
     &\quad \mathcal{L}_{\psi \mathcal{D}^{(*)}\mathcal{D}^{(*)}} \\
       & \quad= -ig_{\psi \mathcal{DD}}\psi^\mu \left( \mathcal{D}^{\dagger}\overleftrightarrow{\partial}_\mu \mathcal{D} \right) \\
       & \quad +g_{\psi \mathcal{D}\mathcal{D}^*}\epsilon^{\mu\nu\alpha\beta}\partial_\mu\psi_\nu \left( \mathcal{D}^{\dagger} \overleftrightarrow{\partial}_\alpha \mathcal{D}^*_{\beta}-\mathcal{D}_\beta^{*\dagger}\overleftrightarrow{\partial}_\alpha \mathcal{D} \right)\\
       & \quad +ig_{\psi \mathcal{D}^*\mathcal{D}^*}\psi^\mu \left( \partial^\nu \mathcal{D}_\mu^{*\dagger} \mathcal{D}^*_\nu-\mathcal{D}_\nu^{*\dagger}\partial^\nu \mathcal{D}_\mu^*+\mathcal{D}^{*\nu\dagger}\overleftrightarrow{\partial}_\mu \mathcal{D}_\nu^* \right),
   \end{split}
 \end{equation}
 \begin{equation}\label{eq:lagrangiand}
   \begin{split}
     &\quad \mathcal{L}_{\psi_1\mathcal{D}^{(*)}\mathcal{D}^{(*)}}  \\
       & \quad= ig_{\psi_1\mathcal{DD}}\psi_1^\mu \mathcal{D}^{\dagger}\overleftrightarrow{\partial}_\mu \mathcal{D}\\
       & \quad -g_{\psi_1\mathcal{DD}^*}\epsilon^{\mu\nu\alpha\beta}\partial_\mu\psi_{1\nu} \left( \mathcal{D}^{\dagger} \overleftrightarrow{\partial}_\alpha \mathcal{D}^*_{\beta}-\mathcal{D}_\beta^{*\dagger}\overleftrightarrow{\partial}_\alpha \mathcal{D} \right)\\
       & \quad +ig_{\psi_1 \mathcal{D}^*\mathcal{D}^*}\psi_1^\mu \left( \partial^\nu \mathcal{D}_\mu^{*\dagger} \mathcal{D}^*_\nu-\mathcal{D}_\nu^{*\dagger}\partial^\nu \mathcal{D}_\mu^*+4\mathcal{D}^{*\nu\dagger}\overleftrightarrow{\partial}_\mu \mathcal{D}_\nu^* \right),
   \end{split}
 \end{equation}
 where $\psi/\psi_1$ is the $S/D$-wave vector charmonium field and $\mathcal{D}^{(*)}$ is the charmed/charmed-strange meson $SU(3)$ triplet $(D^{0(*)},D^{+(*)},D_s^{+(*)})$. For the interaction Lagrangian of $\Lambda_cD^{(*)}N$, we use the following form \cite{Khodjamirian:2011jp}:
 \begin{equation}\label{eq:lagrangianN}
   \begin{split}
     \mathcal{L}_{\Lambda_cD^{(*)}N} &= ig_{\Lambda_cDN}\bar{\Lambda}_c\gamma^5DN \\
       & \quad +\bar{\Lambda}_c\left( g_{\Lambda_cD^*N}\gamma^\mu + \frac{\kappa_{\Lambda_cD^*N}}{m_{\Lambda_c}+m_N}\sigma^{\mu\nu}\partial_\nu \right) D^*_\mu N +h.c.,
   \end{split}
 \end{equation}
 where $N$ denotes nucleon field. The above Lagrangian can be directly applied to the coupling of $\Lambda_cD_s\Lambda$,
 \begin{equation}\label{eq:lagrangianLambda}
   \begin{split}
     \mathcal{L}_{\Lambda_cD_s^{(*)}\Lambda} &= ig_{\Lambda_cD_s\Lambda}\bar{\Lambda}_c\gamma^5D_s\Lambda \\
       & \quad +\bar{\Lambda}_c\left( g_{\Lambda_cD_s^*\Lambda}\gamma^\mu + \frac{\kappa_{\Lambda_cD_s^*\Lambda}}{m_{\Lambda_c}+m_\Lambda}\sigma^{\mu\nu}\partial_\nu \right) D^*_{s\mu} \Lambda + h.c.
   \end{split}
 \end{equation}
 With these effective Lagrangians, the corresponding scattering amplitudes of $\psi\to D^{(*)}_{(s)}\bar{D}^{(*)}_{(s)} \to \Lambda_c\bar{\Lambda}_c$ can be obtained. There are eight kinds of different intermediate meson pair combinations that contribute to the above process, i.e.,  $D\bar{D}$, $D\bar{D}^*$, $D^*\bar{D}$, $D^*\bar{D}^*$ for Fig. \ref{fig:triangle} (a) and $D_s\bar{D}_s$, $D_s\bar{D}^*_s$, $D^*_s\bar{D}_s$, $D^*_s\bar{D}^*_s$ for Fig. \ref{fig:triangle} (b). For $\psi(nS)\to D^{(*)} \bar{D}^{(*)} \to \Lambda_c \bar{\Lambda}_c$, the corresponding four kinds of amplitudes are
 \begin{equation}\label{ampbeg}
   \begin{split}
     \mathcal{M}_a^{\psi} & = i^3 \int \frac{d^4q}{(2\pi)^4} \left(-ig_{\psi DD}\right)\epsilon_\psi^\mu\left(iq_{2\mu}-iq_{1\mu}\right) \\
       & \quad \times \bar{u}(p_1) \left(ig_{\Lambda_cDN}\gamma^5\right)\left(\slashed{q}+m_N\right)\left(ig_{\Lambda_cDN}\gamma^5\right)v(p_2)\\
       & \quad \times \frac{1}{q_1^2-m_D^2} \frac{1}{q_2^2-m_D^2} \frac{1}{q^2-m_N^2} \mathcal{F}^2(q^2),
   \end{split}
 \end{equation}
 \begin{equation}
   \begin{split}
     \mathcal{M}_b^{\psi} & = i^3 \int \frac{d^4q}{(2\pi)^4} g_{\psi DD^*} \epsilon_\psi^\mu \epsilon_{\alpha\beta\lambda\mu} \left(iq_2^\alpha-iq_1^\alpha\right)\left(-ip^\lambda\right) \\
       & \quad \times \bar{u}(p_1)\left(ig_{\Lambda_cDN}\gamma^5\right)\left(\slashed{q}+m_N\right) \\
       & \quad \times \left(g_{\Lambda_cD^*N}\gamma^\rho-i\frac{\kappa_{\Lambda_cD^*N}}{m_{\Lambda_c}+m_N}\sigma^{\rho\nu}q_{2\nu}\right)v(p_2) \\
       & \quad \times \frac{1}{q_1^2-m_D^2} \frac{-g^\beta_\rho+q_2^\beta q_{2\rho}/m_{D^*}^2}{q_2^2-m_{D^*}^2} \frac{1}{q^2-m_N^2} \mathcal{F}^2(q^2),
   \end{split}
 \end{equation}
 \begin{equation}
   \begin{split}
     \mathcal{M}_c^{\psi} & = i^3 \int \frac{d^4q}{(2\pi)^4} g_{\psi DD^*} \epsilon_\psi^\mu \epsilon_{\alpha\beta\lambda\mu} \left(iq_1^\alpha-iq_2^\alpha\right)\left(-ip^\lambda\right) \\
       & \times \quad \bar{u}(p_1)\left(g_{\Lambda_cD^*N}\gamma^\rho-i\frac{\kappa_{\Lambda_cD^*N}}{m_{\Lambda_c}+m_N}\sigma^{\rho\nu}q_{1\nu}\right) \\
       & \times \left(\slashed{q}+m_N\right) \left(ig_{\Lambda_cDN}\gamma^5\right)v(p_2) \\
       & \times \frac{-g^\beta_\rho+q_2^\beta q_{2\rho}/m_{D^*}^2}{q_1^2-m_{D^*}^2} \frac{1}{q_2^2-m_{D}^2} \frac{1}{q^2-m_N^2} \mathcal{F}^2(q^2),
   \end{split}
 \end{equation}
 \begin{equation}
   \begin{split}
     \mathcal{M}_d^{\psi} & = i^3 \int \frac{d^4q}{(2\pi)^4} ig_{\psi D^*D^*}\epsilon_\psi^\mu \left[\left(iq_{2\mu}-iq_{1\mu}\right)g_{\alpha\nu}+iq_{1\alpha}g_{\mu\nu}\right.\\
       & \quad \left.-iq_{2\nu}g_{\mu\alpha}\right] \bar{u}(p_1)\left(g_{\Lambda_cD^*N}\gamma^\lambda-i\frac{\kappa_{\Lambda_cD^*N}}{m_{\Lambda_c}+m_N}\sigma^{\lambda\rho}q_{1\rho}\right)\\
       & \quad \times \left(\slashed{q}+m_N\right) \left(g_{\Lambda_cD^*N}\gamma^\beta-i\frac{\kappa_{\Lambda_cD^*N}}{m_{\Lambda_c}+m_N}\sigma^{\beta\sigma}q_{2\sigma}\right) v(p_2)\\
       & \quad \times \frac{-g^\nu_\lambda+q_1^\nu q_{1\lambda}/m_{D^*}^2}{q_1^2-m_{D^*}^2} \frac{-g^\alpha_\beta+q_2^\alpha q_{2\beta}/m_{D^*}^2}{q_2^2-m_{D^*}^2} \frac{1}{q^2-m_N^2} \mathcal{F}^2(q^2).
   \end{split}
 \end{equation}
 The amplitudes for $\psi(mD)\to D^{(*)} \bar{D}^{(*)} \to \Lambda_c \bar{\Lambda}_c$ are
 \begin{equation}
   \begin{split}
     \mathcal{M}_a^{\psi_1} & = i^3 \int \frac{d^4q}{(2\pi)^4} ig_{\psi DD}\epsilon_\psi^\mu\left(iq_{2\mu}-iq_{1\mu}\right) \\
       & \quad \times \bar{u}(p_1) \left(ig_{\Lambda_cDN}\gamma^5\right)\left(\slashed{q}+m_N\right)\left(ig_{\Lambda_cDN}\gamma^5\right)v(p_2)\\
       & \quad \times \frac{1}{q_1^2-m_D^2} \frac{1}{q_2^2-m_D^2} \frac{1}{q^2-m_N^2} \mathcal{F}^2(q^2),
   \end{split}
 \end{equation}
 \begin{equation}
   \begin{split}
     \mathcal{M}_b^{\psi_1} & = i^3 \int \frac{d^4q}{(2\pi)^4} \left(-g_{\psi DD^*}\right) \epsilon_\psi^\mu \epsilon_{\alpha\beta\lambda\mu} \left(iq_2^\alpha-iq_1^\alpha\right)\left(-ip^\lambda\right) \\
       & \quad \times \bar{u}(p_1)\left(ig_{\Lambda_cDN}\gamma^5\right)\left(\slashed{q}+m_N\right) \\
       & \quad \times \left(g_{\Lambda_cD^*N}\gamma^\rho-i\frac{\kappa_{\Lambda_cD^*N}}{m_{\Lambda_c}+m_N}\sigma^{\rho\nu}q_{2\nu}\right)v(p_2) \\
       & \quad \times \frac{1}{q_1^2-m_D^2} \frac{-g^\beta_\rho+q_2^\beta q_{2\rho}/m_{D^*}^2}{q_2^2-m_{D^*}^2} \frac{1}{q^2-m_N^2} \mathcal{F}^2(q^2),
   \end{split}
 \end{equation}
 \begin{equation}
   \begin{split}
     \mathcal{M}_c^{\psi_1} & = i^3 \int \frac{d^4q}{(2\pi)^4} \left(-g_{\psi DD^*}\right) \epsilon_\psi^\mu \epsilon_{\alpha\beta\lambda\mu} \left(iq_1^\alpha-iq_2^\alpha\right)\left(-ip^\lambda\right) \\
       & \quad \times \bar{u}(p_1)\left(g_{\Lambda_cD^*N}\gamma^\rho-i\frac{\kappa_{\Lambda_cD^*N}}{m_{\Lambda_c}+m_N}\sigma^{\rho\nu}q_{1\nu}\right) \\
       & \quad \times \left(\slashed{q}+m_N\right) \left(ig_{\Lambda_cDN}\gamma^5\right)v(p_2) \\
       & \quad \times \frac{-g^\beta_\rho+q_2^\beta q_{2\rho}/m_{D^*}^2}{q_1^2-m_{D^*}^2} \frac{1}{q_2^2-m_{D}^2} \frac{1}{q^2-m_N^2} \mathcal{F}^2(q^2),
   \end{split}
 \end{equation}
 \begin{equation}\label{ampend}
   \begin{split}
     \mathcal{M}_d^{\psi_1} & = i^3 \int \frac{d^4q}{(2\pi)^4} ig_{\psi D^*D^*}\epsilon_\psi^\mu \left[4\left(iq_{2\mu}-iq_{1\mu}\right)g_{\alpha\nu}+iq_{1\alpha}g_{\mu\nu}\right.\\
       & \quad \left.-iq_{2\nu}g_{\mu\alpha}\right] \bar{u}(p_1)\left(g_{\Lambda_cD^*N}\gamma^\lambda-i\frac{\kappa_{\Lambda_cD^*N}}{m_{\Lambda_c}+m_N}\sigma^{\lambda\rho}q_{1\rho}\right)\\
       & \quad \times \left(\slashed{q}+m_N\right) \left(g_{\Lambda_cD^*N}\gamma^\beta-i\frac{\kappa_{\Lambda_cD^*N}}{m_{\Lambda_c}+m_N}\sigma^{\beta\sigma}q_{2\sigma}\right) v(p_2)\\
       & \quad \times \frac{-g^\nu_\lambda+q_1^\nu q_{1\lambda}/m_{D^*}^2}{q_1^2-m_{D^*}^2} \frac{-g^\alpha_\beta+q_2^\alpha q_{2\beta}/m_{D^*}^2}{q_2^2-m_{D^*}^2} \frac{1}{q^2-m_N^2} \mathcal{F}^2(q^2),
   \end{split}
 \end{equation}
 where the dipole form factor $\mathcal{F}(q^2)$ is introduced to describe off-shell effect of the exchanged baryon in the rescattering process $D^{(*)}\bar{D}^{(*)}\to \Lambda_c\bar{\Lambda}_c$ and avoid the divergence of the loop integral, which has the following form:
 \begin{equation}\label{eqalpha}
   \mathcal{F}(q^2) = \left(\frac{m_E^2-\Lambda^2}{q^2-\Lambda^2}\right)^2, \quad \Lambda = m_E+\alpha \Lambda_{QCD}.
 \end{equation}
Here, $m_E$ and $q$ are the mass and four-momentum of the exchanged baryon, respectively. $\Lambda_{QCD} = 220$ MeV and $\alpha$ is a phenomenological dimensionless parameter. The amplitudes of $\psi\to D_s^{(*)} \bar{D}_s^{(*)} \to \Lambda_c \bar{\Lambda}_c$ can be obtained by replacing $g_{\psi DD}$, $g_{\Lambda_c D N}$, $m_D$, $m_N$ to the corresponding parameters in the $D_s\bar{D}_s$ case.

 The total amplitude of $\psi\to\Lambda_c\bar{\Lambda}_c$ in the hadronic loop mechanism reads as
 \begin{equation}
   \mathcal{M}^{Total} = 2\sum_{i=a,b,c,d}\mathcal{M}_i^q +\sum_{i=a,b,c,d}\mathcal{M}_i^s,
 \end{equation}
 where the factor of 2 in front of $\mathcal{M}_i^q$ comes from the sum over charmed meson isospin doublet $(D^{0(*)},D^{+(*)})$, and $\mathcal{M}_i^s$ is the amplitude for intermediate $D_s^{(*)}\bar{D}_s^{(*)}$ case.
 Then, the branching ratio of $\psi\to\Lambda_c\bar{\Lambda}_c$ can be calculated by
 \begin{equation}
   \mathcal{B}[\psi\to\Lambda_c\bar{\Lambda}_c] = \frac{1}{3}\frac{1}{8\pi} \frac{|\boldsymbol{p}_{\Lambda_c}^{cm}|}{m_{\psi}^2\Gamma_{\psi}}\sum_{pol.}|\mathcal{M}^{Total}|^2,
 \end{equation}
 where the factor of $\frac{1}{3}$ comes from spin average over an initial charmonium state.
The loop integral in Eq. (\ref{ampbeg})-(\ref{ampend}) are evaluated with the help of LoopTools package \cite{Cullen:2011kv,vanOldenborgh:1989wn}, by which both the real and imaginary parts are considered in our calculation.

\section{\label{sec3}Two applications}

In this section, we apply the hadronic loop mechanism discussed in Sec. \ref{sec2} to calculate the branching ratios of the decay $\psi\to\Lambda_c\bar{\Lambda}_c$. We follow our previous work \cite{Wang:2020prx}, in which the spectrum and partial widths to charmed meson pairs for charmonia above $\Lambda_c\bar{\Lambda}_c$ threshold were studied, and further explore $\psi\to\Lambda_c\bar{\Lambda}_c$ decay process for higher vector charmonia here.

\subsection{Reproducing branching ratios of $\psi_{6S-5D}' \to \Lambda_c \bar{\Lambda}_c$ and $\psi_{6S-5D}'' \to \Lambda_c \bar{\Lambda}_c$}

The $Y(4630)$ observed in $e^+e^-\to\Lambda_c\bar{\Lambda}_c$ has attracted much attention, where many explanations was proposed to explain this novel structure. Dai {\it et al.} \cite{Dai:2017fwx,Guo:2010tk} found that the $Y(4630)$ may be treated as $Y(4660)$ in the $\psi'f_0(980)$ bound state picture when taking into account the $\Lambda_c\bar{\Lambda}_c$ final state interaction. Cao {\it et al}. \cite{Cao:2019wwt} found that the enhancement right above the $\Lambda_c\bar{\Lambda}_c$ threshold was well explained by a virtual pole generated by $\Lambda_c\bar{\Lambda}_c$ attractive final state interaction. Cotugno {\it et al.} \cite{Cotugno:2009ys} analyzed the Belle data of $Y(4630)\to\Lambda_c\bar{\Lambda}_c$ and $Y(4630)\to\psi(2S)\pi^+\pi^-$ and found that these two observations are likely to be due to the same state which are very likely to be a charmed baryonium constituted by four quarks. Besides, the charmonium explanation to the $Y(4630)$ was also proposed \cite{Li:2009zu,Wang:2020prx}. Obviously, the present experimental data cannot exclude any possible explanations mentioned above. This situation motivate us to carry out further investigation around this puzzling phenomenon.

Along this line, we focus on the $\Lambda_c\bar{\Lambda}_c$ decays of $\psi_{6S-5D}'$ and $\psi_{6S-5D}''$. In our previous work \cite{Wang:2020prx}, we found that the $Y(4630)$ may be explained as the contribution of $\psi_{6S-5D}'$ and $\psi_{6S-5D}''$, which are the mixture of $\psi(6S)$ and $\psi(5D)$ with mixing angle $\theta=\pm34^\circ$. By reproducing the extracted branching ratios of $\psi_{6S-5D}'\to\Lambda_c\bar{\Lambda}_c$ and $\psi_{6S-5D}''\to\Lambda_c\bar{\Lambda}_c$, the charmonium assignment to the $Y(4630)$ can be tested.

The products of the branching ratio of $\Lambda_c\bar{\Lambda}_c$ channel and the dilepton width of $\psi_{6S-5D}'$ and $\psi_{6S-5D}''$ have been extracted from experimental data in Ref. \cite{Wang:2020prx}, which are listed in Table \ref{tab:alpha}. In general, the dilepton width of charmonium states are of the order of keV \cite{Hagiwara:2002fs}. In this work, we take the dilepton width of these higher charmonia as 1 keV for the rough estimate, and get the branching ratios of $\psi_{6S-5D}' \to \Lambda_c\bar{\Lambda}_c$ and $\psi_{6S-5D}'' \to \Lambda_c\bar{\Lambda}_c$
 \begin{eqnarray}\label{eq:fitbranching}
 \mathcal{B}[\psi_{6S-5D}' \to \Lambda_c\bar{\Lambda}_c]=(2.66\pm 1.2)\times 10^{-3}, \nonumber \\
 \mathcal{B}[\psi_{6S-5D}'' \to \Lambda_c\bar{\Lambda}_c]=(19.0\pm4.6)\times 10^{-3}.
 \end{eqnarray}

\begin{table}
  \begin{ruledtabular}
  \caption{ The second column is the product of the dilepton width and the decay branching ratio of $\Lambda_c\bar{\Lambda}_c$ channel of  $\psi_{6S-5D}'$ and $\psi_{6S-5D}''$ given in Ref. \cite{Wang:2020prx}. The corresponding ranges of $\alpha$ of $\psi_{6S-5D}'$ and $\psi_{6S-5D}''$ for different mixing angles are shown in the last two columns.}\label{tab:alpha}
  \centering
  \begin{tabular}{ccccc}
     &            &                                                                             & \multicolumn{2}{c}{$\alpha$}                           \\
     \cline{4-5}
     & Mass (MeV) & $\Gamma^{e^+e^-} \mathcal{B}[\Lambda_c\bar{\Lambda}_c]$ \cite{Wang:2020prx} & $\theta=-34^\circ$ & $\theta=34^\circ$ \\
    \hline
    $\psi_{6S-5D}'$ & 4585 & 2.66$\pm$1.2 $\mathrm{eV}$ & 1.9$\sim$2.3 & 1.9$\sim$2.3 \\
    $\psi_{6S-5D}''$ & 4675 & 19.0$\pm$4.6 $\mathrm{eV}$ & 3.9$\sim$4.3 & 5.4$\sim$6.1 \\
  \end{tabular}
  \end{ruledtabular}
\end{table}

Before showing the numerical results, we introduce how to fix the values of the relevant coupling constants appearing in Eqs. (\ref{eq:lagrangians})-(\ref{eq:lagrangianLambda}). {The Lagrangian $\mathcal{L}_S$ in Eq. (\ref{eq:lagrangians}) can be applied to describe the interaction between $\psi_{6S-5D}'$ and charmed meson pairs, and the $\mathcal{L}_D$ in Eq. (\ref{eq:lagrangiand}) is used for the $\psi_{6S-5D}''$ case.}
The coupling constants in the $\psi \mathcal{D^{(*)}D^{(*)}}$ coupling can be determined by theoretical partial decay widths of $\psi \to D^{(*)}\bar{D}^{(*)}$, which was given in Ref. \cite{Wang:2020prx}. Here, the corresponding partial widths and the extracted coupling constants are listed in Table \ref{tab:6S5Dcoupling}.

 \begin{table*}
 \begin{ruledtabular}
   \caption{Partial widths of $\psi_{6S-5D}' \to D^{(*)}_{(s)}D^{(*)}_{(s)}$ and $\psi_{6S-5D}'' \to D^{(*)}_{(s)}D^{(*)}_{(s)}$  predicted in Ref. \cite{Wang:2020prx} and the corresponding coupling constants $g_{\psi D^{(*)}_{(s)}D^{(*)}_{(s)}}$.  The relative minus sign in the couplings of $\psi_{6S-5D}''$ can be determined in the heavy quark limit \cite{Huang:2018cco}.}\label{tab:6S5Dcoupling}
   \centering
   \begin{tabular}{ccccccccc}
    & \multicolumn{4}{c}{Negative mixing scheme} & \multicolumn{4}{c}{Positive mixing scheme} \\
    \cline{2-5} \cline{6-9}
    & \multicolumn{2}{c}{Partial width (MeV)} & \multicolumn{2}{c}{Coupling constants} & \multicolumn{2}{c}{Partial width (MeV)} & \multicolumn{2}{c}{Coupling constants} \\
    Final state& $\psi_{6S-5D}'$ & $\psi_{6S-5D}''$ & $\psi_{6S-5D}'$ & $\psi_{6S-5D}''$ & $\psi_{6S-5D}'$ & $\psi_{6S-5D}''$ & $\psi_{6S-5D}'$ & $\psi_{6S-5D}''$ \\
    \hline
    $DD$ & 0.17 & 6.15 & 0.120 & $-$0.677 & 4.61 & 1.68 & 0.625 & $-$0.354 \\
    $DD^*$ & 0.7 & 0.67 & 0.030 $\mathrm{GeV}^{-1}$ & 0.027 $\mathrm{GeV}^{-1}$ & 0.24 & 1.17 & 0.018 $\mathrm{GeV}^{-1}$ & 0.035 $\mathrm{GeV}^{-1}$  \\
    $D^*D^*$ & 4.74 & 6.04 & 0.289 & 0.115 & 5.30 & 5.80 & 0.306 & 0.113 \\
    $D_sD_s$ & 0.01 & 0.02 & 0.035 & $-$0.045 & 0.01 & 0.02 & 0.035 & $-$0.045 \\
    $D_sD_s^*$ & 0.19 & $10^{-4}$ & 0.020 $\mathrm{GeV}^{-1}$ & $4\times 10^{-4}$ $ \mathrm{GeV}^{-1}$ & 0.03 & 0.16 & 0.008 $\mathrm{GeV}^{-1}$ & 0.016 $\mathrm{GeV}^{-1}$ \\
    $D_s^*D_s^*$ & 0.11 & 0.21 & 0.063 & 0.030 & 0.25 & 0.06 & 0.095 & 0.016 \\
   \end{tabular}
 \end{ruledtabular}
 \end{table*}

 \begin{table*}
 \begin{ruledtabular}
   \centering
   \caption{ The coupling constants $g_{\psi D^{(*)}_{(s)}D^{(*)}_{(s)}}$ for $\psi(nS)\;(n=7,8,9)$ and $\psi(mD)\;(m=6,7,8)$, which are converted by the corresponding partial widths calculated in Ref. \cite{Wang:2020prx}. }\label{tab:mesoncoupling}
   \begin{tabular}{ccccccc}
    Couplings & $\psi(7S)$ & $\psi(8S)$ & $\psi(9S)$ & $\psi(6D)$ & $\psi(7D)$ & $\psi(8D)$ \\
     \hline
    $DD$ & 0.281 & 0.226 & 0.180 & $-0.449$ & $-0.355$ & $-0.276$ \\
    $DD^*$ & 0.022 $\mathrm{GeV}^{-1}$ & 0.019 $\mathrm{GeV}^{-1}$ & 0.016 $\mathrm{GeV}^{-1}$ & 0.023 $\mathrm{GeV}^{-1}$ & 0.019 $\mathrm{GeV}^{-1}$ & 0.014 $\mathrm{GeV}^{-1}$ \\
    $D^*D^*$ & 0.128 & 0.080 & 0.053 & 0.076 & 0.060 & 0.046 \\
    $D_sD_s$ & 0.010 & 0.009 & 0.009 & $-0.030$ & $-0.028$ & $-0.027$ \\
    $D_sD_s^*$ & 0.006 $\mathrm{GeV}^{-1}$ & 0.006 $\mathrm{GeV}^{-1}$ & 0.003 $\mathrm{GeV}^{-1}$ & 0.007 $\mathrm{GeV}^{-1}$ & 0.005 $\mathrm{GeV}^{-1}$ & 0.003 $\mathrm{GeV}^{-1}$ \\
    $D_s^*D_s^*$ & 0.050 & 0.050 & 0.037 & 0.013 & 0.009 & 0.007 \\
   \end{tabular}
   \end{ruledtabular}
 \end{table*}
 
 For the coupling constants involved in the $\Lambda_cD^{(*)}N$ interaction, we take the values from the calculation of QCD light-cone sum rules \cite{Khodjamirian:2011jp}, i.e., 
  $g_{\Lambda_cDN} = 13.8$, $g_{\Lambda_cD^*N} = -7.9$, and $\kappa_{\Lambda_cD^*N} = 4.7$.
The coupling constants $g_{\Lambda_cD^{(*)}N}$ can be directly related to the coupling constants in $\Lambda_cD_s^{(*)}\Lambda$ interaction under $SU(3)$ symmetry: $g_{\Lambda_cD^{(*)}N} = -\sqrt{\frac{3}{2}} g_{\Lambda_cD^{(*)}_s\Lambda}$.
After fixing all coupling constants, the only free parameter $\alpha$ is left, which is introduced in Eq. (\ref{eqalpha}) to parameterize the cutoff $\Lambda$ in the form factor $\mathcal{F}(q^2)$. Since the cutoff should not deviate far from the physical mass of the exchanged particle, $\alpha$ is expected to be the order of unity as indicated in Ref. \cite{Cheng:2004ru}. 

\begin{figure}[htbp]
  \centering
  \includegraphics[width=8cm,keepaspectratio]{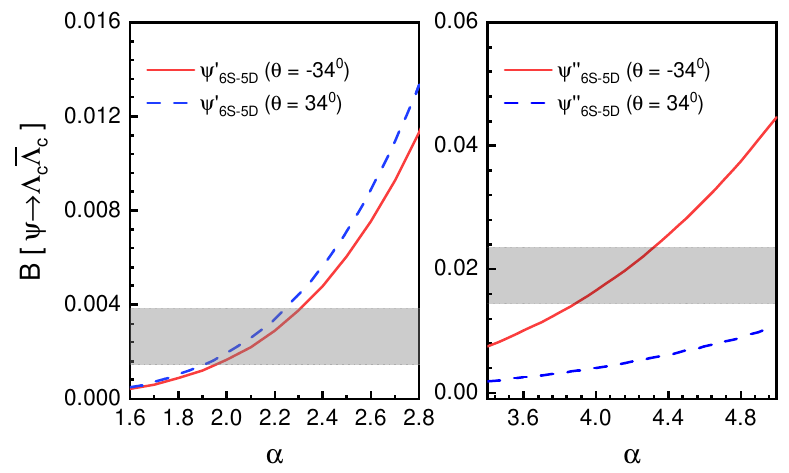}
  \caption{The $\alpha$ dependence on the calculated branching ratios of $\psi_{6S-5D}' \to \Lambda_c\bar{\Lambda}_c$ and $\psi_{6S-5D}'' \to \Lambda_c\bar{\Lambda}_c$. Here, we take positive and negative values of mixing angle $\theta$ to present the decay behavior of $\psi_{6S-5D}'$ and $\psi_{6S-5D}''$. The gray band represents the extracted branching ratio by the experimental data of $e^+e^- \to \Lambda_c\bar{\Lambda}_c$ \cite{Wang:2020prx}.  }\label{fig:6s5d}
\end{figure}

In Fig. \ref{fig:6s5d}, we show the $\alpha$ dependence of the branching ratios of the decay $\psi_{6S-5D}'\to\Lambda_c\bar{\Lambda}_c$ and $\psi_{6S-5D}''\to\Lambda_c\bar{\Lambda}_c$. Here, the results under taking two possible mixing angles are given since the sign of mixing angle $\theta$ was not determined \cite{Wang:2020prx}. For the purpose of comparison, we also show the extracted branching ratios in Fig. \ref{fig:6s5d} as shadow regions. Indeed, the corresponding experimental data can be reproduced well. For the case of $\psi_{6S-5D}'$, when $\alpha$ is taken as the range of $1.9\sim 2.3$, the results with positive and negative mixing angles are similar to each other, where we get the branching ratio consistent with 
the extracted value. For the case of $\psi_{6S-5D}''$, there exists big difference for the result under two mixing angles, where taking positive mixing angle results in a larger $\alpha$ value compared with the case of taking negative mixing angle. Thus, the negative mixing scheme is more preferred. In the following discussion, we take the negative mixing angle, where the extracted branching ratio of $\psi_{6S-5D}''\to\Lambda_c\bar{\Lambda}_c$ decay can be well reproduced when $\alpha=3.9\sim4.3$. In fact, these values of $\alpha$ are not far away from unity and can be seen to be reasonable. 

{In conclusion}, by the above study of the $\Lambda_c\bar{\Lambda}_c$ decay $\psi_{6S-5D}'$ and $\psi_{6S-5D}''$ , we find that the the charmonium explanation for the $Y(4630)$ \cite{Wang:2020prx} can be tested. 

\subsection{Predicting the $\Lambda_c \bar{\Lambda}_c$ decay properties of higher charmonia above 4.7 GeV}

\begin{figure}[htbp]
  \centering
  \includegraphics[width=8cm,keepaspectratio]{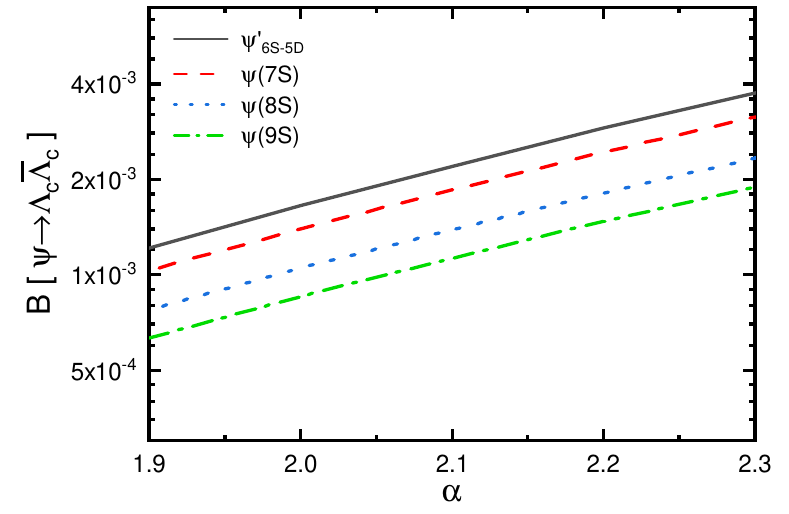}
  \caption{The $\alpha$ dependence of the predicted branching ratios of the decay of higher $S$-wave charmonia to $\Lambda_c\bar{\Lambda}_c$.}\label{fig:branchings}
\end{figure}

\begin{figure}[htbp]
  \centering
  \includegraphics[width=8cm,keepaspectratio]{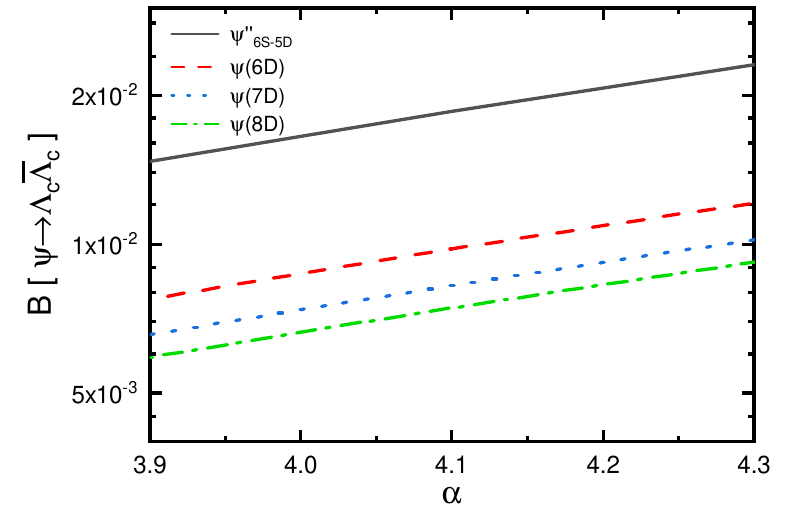}
  \caption{The $\alpha$ dependence of the predicted branching ratios of the decay of higher $D$-wave charmonia to $\Lambda_c\bar{\Lambda}_c$.}\label{fig:branchingd}
\end{figure}

In this subsection, we predict the branching ratios of the $\psi\to\Lambda_c\bar{\Lambda}_c$ decay for higher charmonia,
where we focus on three $\psi(nS)$ $(n=7,8,9)$ and three $\psi(mD)$ $(m=6,7,8)$ which were predicted  \cite{Wang:2020prx} to have the mass in the energy region between 4.7 and 4.9 GeV. Similarly, the $\psi \mathcal{D^{(*)}D^{(*)}}$ coupling constants for these higher charmonia can also be fixed by the theoretically evaluated partial decay widths which are summarized in Table \ref{tab:mesoncoupling}. 
The $\alpha$ dependence of the calculated branching ratios of $\Lambda_c\bar{\Lambda}_c$ channel by the hadronic loop mechanism are shown in Fig. \ref{fig:branchings} and Fig. \ref{fig:branchingd}.


As shown in Fig. \ref{fig:branchings} and Fig. \ref{fig:branchingd}, the branching ratios of $\psi(nS)\to\Lambda_c\bar{\Lambda}_c$ and $\psi(mD)\to\Lambda_c\bar{\Lambda}_c$ have two features, i.e.,
\begin{enumerate}
    \item	The branching ratios for $\psi(nS)\to\Lambda_c\bar{\Lambda}_c$ are of the order of magnitude of $10^{-4}-10^{-3}$, which are similar to the case of $\psi_{6S-5D}'$. The branching ratios for $\psi(mD)\to\Lambda_c\bar{\Lambda}_c$ are of the order of $10^{-3}-10^{-2}$, which are similar to the case of $\psi_{6S-5D}''$.
    \item	The ratios $\mathcal{B}[\psi(nS) \to \Lambda_c\bar{\Lambda}_c]$/$\mathcal{B}[\psi_{6S-5D}' \to \Lambda_c\bar{\Lambda}_c]$ are stable with the change of $\alpha$ value. Similarly, the ratios between $\mathcal{B}[\psi(mD) \to \Lambda_c\bar{\Lambda}_c]$ and $\mathcal{B}[\psi_{6S-5D}'' \to \Lambda_c\bar{\Lambda}_c]$ are almost fixed when varying the $\alpha$ values.
\end{enumerate}
Here, we list the concrete ratios of $\frac{\mathcal{B}[\psi(nS) \to \Lambda_c\bar{\Lambda}_c]}{\mathcal{B}[\psi_{6S-5D}' \to \Lambda_c\bar{\Lambda}_c]}$ and $\frac{\mathcal{B}[\psi(mD) \to \Lambda_c\bar{\Lambda}_c]}{\mathcal{B}[\psi_{6S-5D}'' \to \Lambda_c\bar{\Lambda}_c]}$, which are
\begin{equation}\label{eq:ratio}
  \begin{split}
    \mathcal{B}[9S]:\mathcal{B}[8S]:\mathcal{B}[7S]:\mathcal{B}[\psi'_{6S-5D}] & = 0.51:0.63:0.84:1 ,\\
    \mathcal{B}[8D]:\mathcal{B}[7D]:\mathcal{B}[6D]:\mathcal{B}[\psi''_{6S-5D}] & = 0.40:0.45:0.53:1 .
  \end{split}
\end{equation}
These features are understandable under the hadronic loop mechanism. For initial states with the same internal orbital angular momentum $L$, the dynamical difference of their decay behavior within hadronic loop mechanism comes from $\psi \mathcal{D^{(*)}D^{(*)}}$ coupling vertexes and these couplings are determined by their partial decay widths to charmed meson pairs. In general, the amplitude for a two-body decay involves an overlap integral among three wave functions. The wave functions of highly radially excited charmonium states, as we consider here, highly oscillate and the overlap integrals have similar behavior for these charmonium states as indicated in Ref. \cite{Wang:2020prx}.


\section{\label{sec4}Discussion and conclusion}

Many charmonia above $D\bar{D}$ have been established with the accumulation of experimental data in the past decades \cite{Hagiwara:2002fs}. Specifically, the observations of $Y(4660)$ in $e^+e^-\to\psi(2S)\pi^+\pi^-$ \cite{Wang:2014hta} and $Y(4630)$ in $e^+e^- \to \Lambda_c\bar{\Lambda}_c$ \cite{Pakhlova:2008vn} stimulated the construction of higher vector charmonia above the $\Lambda_c\bar{\Lambda}_c$ threshold \cite{Li:2009zu,Ding:2007rg,Wang:2020prx}. When studying these relevant problems, a key problem is how to quantitatively calculate the $\Lambda_c\bar{\Lambda}_c$ decays of higher vector charmonia. However, the decay $\psi\to\Lambda_c\bar{\Lambda}_c$ was poorly investigated in the past theoretical researches \cite{Xiao:2018iez,Simonov:2011jc,Simonov:2011cm}.

 In this work, we investigated the decay mechanism of $\psi\to\Lambda_c\bar{\Lambda}_c$ by introducing the hadronic loop mechanism. In this mechanism, a charmonium state firstly decays into a charmed meson pair $D^{(*)}\bar{D}^{(*)}$ or a charmed-strange meson pair $D^{(*)}_s\bar{D}^{(*)}_s$, and then the intermediate charmed or charmed-strange meson pair is transformed into final states $\Lambda_c\bar{\Lambda}_c$ by exchanging an intermediate baryon state.
 
As an application of the hadronic loop mechanism, we have examined the decay behaviors of charmonium states $\psi_{6S-5D}'$ and $\psi_{6S-5D}''$ to $\Lambda_c\bar{\Lambda}_c$, where the $\psi_{6S-5D}'$ and $\psi_{6S-5D}''$ are the mixture of $\psi(6S)$ and $\psi(5D)$ predicted in Ref. \cite{Wang:2020prx}. We found that the calculated branching ratios of $\psi_{6S-5D}' \to \Lambda_c\bar{\Lambda}_c$ and $\psi_{6S-5D}''\to \Lambda_c\bar{\Lambda}_c$ can match the corresponding extracted values from the fit to the experimental data of  $e^+e^-\to \Lambda_c\bar{\Lambda}_c$ in Ref. \cite{Wang:2020prx}, where the $Y(4630)$ structure was reproduced by the contribution of 
$\psi_{6S-5D}'$ and $\psi_{6S-5D}''$ \cite{Wang:2020prx}. Thus, by the study of  $\psi_{6S-5D}^{'/''} \to \Lambda_c\bar{\Lambda}_c$, the charmonium explanation to the $Y(4630)$ is tested in the present work. 

Furthermore, the branching ratios of the $\Lambda_c\bar{\Lambda}_c$ decay mode of higher vector charmonium states up to 4.9 GeV were also explored. The branching ratios for the $\Lambda_c\bar{\Lambda}_c$ decays of $S$-wave charmonia are found to be of the order of magnitude of $10^{-4}-10^{-3}$ and those of the corresponding $D$-wave charmonium partners are of the order of magnitude of $10^{-3}-10^{-2}$. The branching ratios of $\mathcal{B}[\psi(nS)/\psi(mD)\to\Lambda_c\bar{\Lambda}_c]$ with $n=7,8,9$ and $m=6,7,8$ are comparable with those of $\psi'_{6S-5D}/\psi''_{6S-5D}$. 
 
Since the decay of these higher charmonia to $\Lambda_c\bar{\Lambda}_c$ have sizable branching ratios, before ending this article, we discuss the possibility of finding out these missing charmonia in $e^+e^-\to\Lambda_c\bar{\Lambda}_c$. In the following, we try to mimic the corresponding cross section by combining with our the present study. 
 If considering the intermediate charmonium contribution to $e^+e^-\to  \Lambda_c\bar{\Lambda}_c$, a phase space corrected Breit-Wigner distribution reads as
\begin{equation}
  \mathcal{M}_R(\psi) = \frac{\sqrt{12\pi\Gamma^{e^+e^-}_{\psi} \mathcal{B}[\psi\to\Lambda_c\bar{\Lambda}_c]\Gamma_{\psi}}}{s-m^2_{\psi}+im_{\psi}\Gamma_{\psi}} \sqrt{\frac{\Phi(s)}{\Phi(m_{\psi}^2)}},
\end{equation}
where $\psi$ denotes the intermediate charmonium resonance and $\Phi(s)$ is the phase space. Additionally, we define a free parameter $\mathcal{R}_{\psi}=\Gamma^{e^+e^-}_{\psi}\mathcal{B}[\psi\to\Lambda_c\bar{\Lambda}_c]$. Additionally, a non-resonance contribution is parameterized as \cite{Wang:2020prx}
\begin{equation}
  \mathcal{M} = g_{NoR}(\sqrt{s}-2m_{\Lambda_c})^{\frac{1}{2}} e^{-(a\sqrt{s}+bs)},
\end{equation}
where $g_{NoR}$, $a$, and $b$ are free parameters. The total amplitude of the process $e^+e^-\to\Lambda_c\bar{\Lambda}_c$ can be written as
\begin{equation}
  \mathcal{M}^{Total} = \mathcal{M}_{NoR}+\sum_i e^{i \phi_i}\mathcal{M}_{R_i},
\end{equation}
where $\phi_i$ is the phase angles between the $i$-th resonance amplitude and non-resonance term.

In this analysis, the masses and widths of the involved charmonium states are taken to be theoretical values from the screened potential model \cite{Wang:2020prx}. Because a similar fit was performed in Ref. \cite{Wang:2020prx} to extract the branching ratio of $\psi_{6S-5D}'$ and $\psi_{6S-5D}''$, so we choose the same parameters $g_{NoR}$, $a$, $b$, and $\mathcal{R}_{\psi_{6S-5D}'}$ as those in Ref. \cite{Wang:2020prx}. Here, the $\mathcal{R}_{\psi_{6S-5D}''}$ is considered as a free parameter in our present fit because the fitted width of $\psi_{6S-5D}''$ in Ref. \cite{Wang:2020prx} is almost three times larger than our theoretical estimation and we argued that this inconsistency may be due to the influence from possible $\psi(7S)$. Finally, other free parameters in the fit are the relative phases associated with various charmonium resonances.

\begin{figure}[htbp]
  \centering
  \includegraphics[width=8cm,keepaspectratio]{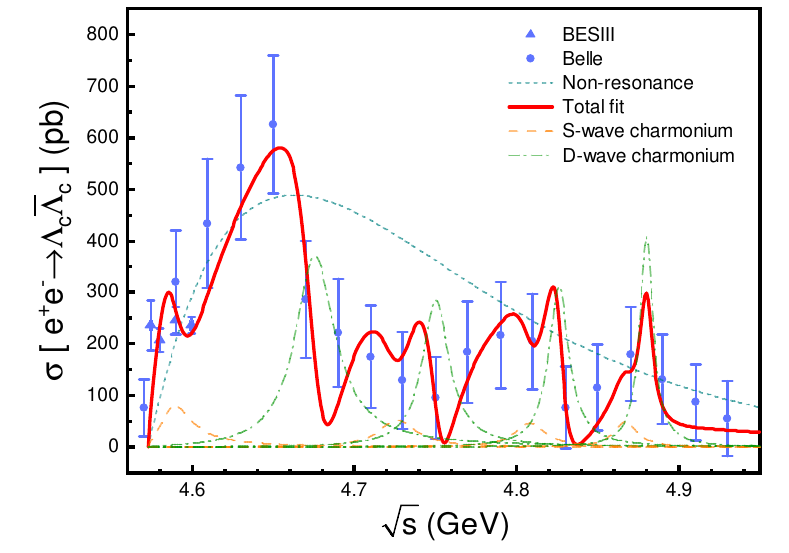}
  \caption{Our fit result to experimental data of $e^+e^-\to\Lambda_c \bar{\Lambda}_c$ from Belle \cite{Pakhlova:2008vn} and BESIII \cite{Ablikim:2017lct}.}\label{fig:fit}
\end{figure}

\begin{table*}
\begin{ruledtabular}
  \centering
  \caption{The fitted parameters to $e^+e^-\to\Lambda_c\bar{\Lambda}_c$ from Belle \cite{Pakhlova:2008vn} and BESIII \cite{Ablikim:2017lct}.}\label{tab:fitparameter}
  \begin{tabular}{c|ccccccccccccccccccc}
     Parameter & $\mathcal{R}_{\psi_{6S-5D}''}$ & $\phi_{\psi_{6S-5D}''}$ & $\phi_{7S}$ & $\phi_{6D}$ & $\phi_{8S}$ & $\phi_{7D}$ & $\phi_{9S}$ & $\phi_{8D}$  \\
     Unit & eV & rad & rad & rad & rad & rad & rad & rad \\
     \hline
     Value & 16.5 & 4.24 & 4.61 & 3.90 & 3.80 & 3.28 & 2.97 & 4.28 \\
     Error($\pm$) & 5.7 & 0.21 & 0.51 & 0.27 & 0.46 & 0.45 & 0.88 & 0.74 \\
   \end{tabular}
\end{ruledtabular}
\end{table*}

The fitted parameters are listed in Table \ref{tab:fitparameter} and the fitted result is shown in Fig. \ref{fig:fit}, where the $\chi^2/d.o.f = 1.465$ is obtained. 
Similar to the $Y(4630)$ structure, one can see that the interference of adjacent $\psi(nS)$ and $\psi((n-1)D)$ really shows several obvious enhancements in the energy region between 4.7 and 4.9 GeV in Fig. \ref{fig:fit}. Unfortunately, the uncertainties of the Belle data are too large to draw any solid conclusions from the present fit. However, it is worth noting that the respective contributions of these higher charmonium states to the cross section of $e^+e^- \to \Lambda_c\bar{\Lambda}_c$ are also shown in Fig. \ref{fig:fit}, which are directly calculated from the predicted branching ratios in Eq. (\ref{eq:ratio}) and are independent of the fit schemes. Thus, it should provide some interesting evidences for the existence of these higher charmonia in this channel. We are hopeful for more precise experimental measurements to clearly confirm these local enhancements. We notice that the BESIII Collaboration has recently released their white paper on the future physics program \cite{Ablikim:2019hff}. At present, the BEPCII is going to take data in center-of-mass energy region between 4.6 and 4.9 GeV and the data set corresponding to 15 $\text{fb}^{-1}$ of total integrated luminosity is expected. It is interesting to test the property of dense charmonium spectrum above 4.6 GeV, and the future BESIII and upcoming Belle II will provide a good platform to search for these charmonium states.

\vfil
\section*{Acknowledgements}

This work is supported by the China National Funds for Distinguished Young Scientists under Grant No. 11825503, National Key
Research and Development Program of China under Contract No. 2020YFA0406400, and the 111 Project under Grant No. B20063, the National Natural Science Foundation of China under Grant No. 12047501, and by the Fundamental Research Funds for the Central Universities.


\end{document}